\documentclass[lettersize,journal]{IEEEtran}
\usepackage{amsmath,amsfonts}
\usepackage{algorithmic}
\usepackage{algorithm}
\usepackage{array}
\usepackage[caption=false,font=normalsize,labelfont=sf,textfont=sf]{subfig}
\usepackage{textcomp}
\usepackage{stfloats}
\usepackage{url}
\usepackage{verbatim}
\usepackage{graphicx}
\usepackage{cite}
\usepackage{multirow}  
\usepackage{xspace}
\usepackage{xcolor}

\newcommand{\dg}{\texttt{DGVAE}}
\newcommand{\ie}{\emph{i.e.},\xspace}
\newcommand{\eg}{\emph{e.g.},\xspace}
\newcommand{\bm}[1]{\mathbf{#1}}

\begin{document}

\title{Disentangled Graph Variational Auto-Encoder for Multimodal Recommendation with Interpretability}

\author{Xin Zhou and Chunyan Miao,~\IEEEmembership{Fellow~IEEE}
\thanks{Xin Zhou is with the Alibaba-NTU Singapore Joint Research Institute, Nanyang Technological University, Singapore. Chunyan Miao is with the School of Computer Science and Engineering, Nanyang Technological University, Singapore.
(e-mail: xin.zhou@ntu.edu.sg; ASCYMiao@ntu.edu.sg).}
}

\markboth{Journal of \LaTeX\ Class Files,~Vol.~14, No.~8, August~2021}%
{Shell \MakeLowercase{\textit{et al.}}: A Sample Article Using IEEEtran.cls for IEEE Journals}


\maketitle

\begin{abstract}
Multimodal recommender systems amalgamate multimodal information (\eg textual descriptions, images) into a collaborative filtering framework to provide more accurate recommendations. While the incorporation of multimodal information could enhance the interpretability of these systems, current multimodal models represent users and items utilizing entangled numerical vectors, rendering them arduous to interpret.
To address this, we propose a Disentangled Graph Variational Auto-Encoder (\dg{}) that aims to enhance both model and recommendation interpretability. \dg{} initially projects multimodal information into textual contents, such as converting images to text, by harnessing state-of-the-art multimodal pre-training technologies. It then constructs a frozen item-item graph and encodes the contents and interactions into two sets of disentangled representations utilizing a simplified residual graph convolutional network. \dg{} further regularizes these disentangled representations through mutual information maximization, aligning the representations derived from the interactions between users and items with those learned from textual content.
This alignment facilitates the interpretation of user binary interactions via text. Our empirical analysis conducted on three real-world datasets demonstrates that \dg{} significantly surpasses the performance of state-of-the-art baselines by a margin of 10.02\%.
We also furnish a case study from a real-world dataset to illustrate the interpretability of \dg{}. Code is available at: \url{https://github.com/enoche/DGVAE}.
\end{abstract}


\section{Introduction}
\IEEEPARstart{C}{ollaborative} filtering (CF) serves as a preeminent strategy for facilitating the identification of items of interest to users on e-commerce platforms~\cite{koren2008factorization, jing2023capturing, zhou2023selfcf}.
However, the {increasing prevalence} of multimedia information (\eg images, texts, and videos) on these platforms presents new challenges to the current CF paradigm. One such challenge {involves} the effective utilization of multimodal information and its successful integration into recommendation tasks. Another challenge, we argue, is the unsatisfactory interpretability of current multimodal recommendations when incorporating multimodal information.

To effectively incorporate multimodal information into the existing CF framework, conventional methodologies have {employed} techniques that concatenate or sum up the pre-processed multimodal information with learnable item embeddings~\cite{he2016vbpr, liu2017deepstyle}.
{More advanced} recommendation approaches have even utilized attention mechanisms to accurately capture users’ preferences for items~\cite{chen2017attentive, liu2019user, chen2019personalized}.
These approaches {facilitate} a more nuanced understanding of user behavior and preferences, leading to more accurate and personalized recommendations.
The recent increase in research on graph-based recommendations~\cite{he2020lightgcn, wu2022graph, zhang2022diffusion, zhou2023layer} has inspired a new line of work~\cite{wei2019mmgcn, wei2020graph, wang2021dualgnn, zhang2021mining, zhou2023enhancing} that captures the high-order relationships between item multimodal features and user-item interactions leveraging the power of graph neural networks (GNNs).
Specifically, MMGCN~\cite{wei2019mmgcn} adopts graph convolutional networks (GCNs) to propagate and aggregate information on every modality of the item.
{Building upon} MMGCN, GRCN~\cite{wei2020graph} denoises the dyadic relations present in the interaction data by identifying noise user-item edges based on the affinity score of user preference and item content.
Other authors in~\cite{yi2021multi} leverage variational auto-encoders (VAEs) on graphs (VGAE~\cite{kipf2016variational}) to generate modality-specific numeric embeddings and fuse them for recommendation.
{To effectively utilize} multimodal features in recommendation, researchers have even developed approaches that complement the user-item interaction graph by exploiting auxiliary graph structures.
For instance, DualGNN~\cite{wang2021dualgnn} introduces a user concurrence graph to capture user preferences for modal features. 
LATTICE~\cite{zhang2021mining} posits that the latent structures underlying the multimodal contents of items can {enhance} representation learning.
{Consequently}, it introduces an item-item auxiliary graph based on modality information and performs graph convolutions on {this} graph.

Although graph-based multimodal models {demonstrate} promising performance in recommendation {tasks}, {their reliance on numerical embeddings for representing users or items poses a challenge to the interpretability of the models and their recommendations.}
This lack of interpretability further hampers the application of these models {in} industrial scenarios.
Interpretability is a crucial aspect of recommender systems that facilitates the comprehension of the system’s output by its users. It fosters trustworthiness and confidence in the decision-making process, {thereby enhancing} the overall utility of the system~\cite{tintarev2007explanations, caro2023graph, zhang2020explainable}.

In this paper, we address the {aforementioned} issues by extending VAEs to graph structures with a novel graph VAE paradigm without compromising recommendation performance.
While VAEs have been applied to graphs in~\cite{yi2021multi, kipf2016variational}, they perform graph convolutions on user-item bipartite graphs and represent hidden layer embeddings as mean and standard deviation. We empirically and theoretically show that this learning paradigm in multimodal recommendation may {degrade} the learning of user and item representations, {as discussed} in Section~\ref{sec:exp}.
In contrast, our proposed \dg{} fully exploits multimodal information in two {distinct} ways.
{Firstly}, we convert raw information (\eg images, text) of items from each modality into numerical embeddings using pre-trained Transformers.
Based on {these} embeddings, we construct an item-item graph that explicitly models the relations between items via multimodal perspectives.
{Secondly}, we align raw information of items from other modalities with the textual content by harnessing pre-trained multimodal models. 
With the textual contents, we frame user preferences into a user-word matrix, {where each} element in the matrix denotes the preference of a word for a given user.
Our proposed model performs graph convolutions on the item-item graph to encode disentangled representations by reconstructing both the user-item interaction matrix and the user-word preference matrix.
{Finally, the proposed model regularizes the two sets of disentangled representations by maximizing their mutual information.}
{We summarize the main contributions of this paper as follows:
\begin{itemize}
\item To enhance the interpretability of multimodal recommender systems, we propose \dg{} that projects multimodal information into textual contents and encodes the projected textual content, as well as user ratings, into two distinct sets of disentangled representations utilizing graph convolutional networks.
\item To interpret users' numerical interactions with their preferred words, \dg{} regularizes the two sets of representations by maximizing their mutual information. This regularization ensures that the representations derived from the interactions between users and items align with those learned from the textual content.
\item We conduct extensive experiments on three real-world datasets to validate the efficacy of our proposed model. The results demonstrate that our model significantly surpasses state-of-the-art methods in terms of recommendation accuracy. Furthermore, we delve into the recommendation results and elucidate the interpretability of \dg{} through a user case study conducted on a real-world dataset.
\end{itemize}
}

\section{Related Work}
\label{sec:relatedwork}
\subsection{Multimodal Recommendation}
Multimodal recommendation models improve the performance of classic CF models by incorporating the multimodal information of items with deep learning or graph learning techniques~\cite{zhou2023comprehensive}.
Early work, such as VBPR~\cite{he2016vbpr} and DeepStyle~\cite{liu2017deepstyle}, combines item ID embeddings and visual features for matrix decomposition.
In light of the advancements in Transformer architectures, VECF~\cite{chen2019personalized} leverages attention mechanisms to effectively capture complex user preferences on image patches. 
{With} the burgeoning adoption of Graph Neural Networks (GNNs) within the domain of recommendation systems~\cite{wu2022graph}, researchers have been {motivated} to assimilate high-order semantic data into the methodology of acquiring user and item representations through the implementation of GNNs.
MMGCN~\cite{wei2019mmgcn} employs graph convolution techniques to each individual modality, subsequently aggregating the results through a process of modal fusion. 
GRCN~\cite{wei2020graph} implements a refinement process on the bipartite user-item graph by identifying and eliminating noise edges for effective recommendation.
DualGNN~\cite{wang2021dualgnn} introduces a user co-occurrence graph in conjunction with a preference learning module, designed to accommodate the dynamic evolution of users' preferences over time. 
LATTICE~\cite{zhang2021mining} posits that previous approaches to extracting semantic information between items have relied on implicit methods, which may result in suboptimal performance. To address this issue, LATTICE explicitly learns and constructs item-item relation graphs for each individual modality, subsequently fusing them together to obtain a latent item-item graph. 
By further freezing the item-item graph, FREEDOM~\cite{zhou2023tale} significantly enhances the performance of LATTICE's recommendation results, achieving an impressive boost of 19.07\%.
DMRL~\cite{liu2022disentangled} uses a multimodal attention mechanism to capture users' attention on each factor of different modalities.
MVGAE~\cite{yi2021multi} is the first work that incorporates VGAE to learn the numeric embeddings for uni-modal information. It then fuses the learned embeddings and ID embeddings with {the} product of experts. 
Recently, we also see an emerging of applying self-supervised learning in multimodal recommendations~\cite{tao2022self, zhou2023bootstrap}.
Specifically, SLMRec~\cite{tao2022self} employs a self-supervised learning approach within a graph neural network framework to effectively capture the underlying relationships between interactions. On the other hand, BM3~\cite{zhou2023bootstrap} introduces a novel self-supervised learning methodology that addresses the challenges of high computational cost and inaccurate supervision signals. 
Both approaches demonstrate the potential of self-supervised learning in improving the performance of recommendation systems.
Although existing multimodal recommendation models show effective performance in recommendation, they all utilize the pre-trained numeric embeddings from each modality for model training and inference, thus {lacking} in interpretability.

\subsection{Variational Auto-Encoders}
As a Bayesian version of auto-encoders, VAEs have {proven} to be effective in CF~\cite{liang2018variational, askari2021variational}.
MultiVAE~\cite{liang2018variational} extends VAEs to CF by designing a generative model that captures users’ preferences.
Following {this} work, CVAE~\cite{li2017collaborative} first incorporates both implicit feedback and content of items into a unified CF paradigm.
Specifically, the model learns two latent variables from both contents of the item and its ratings via matrix factorization. It then sums up the two variables to represent the latent representation of the item.
MD-CVAE~\cite{zhu2022mutually} empowers CVAE with a MultiVAE encoder and couples the two variables learned from contents and ratings with regularization.
{Another line of work aims to} disentangle the representations of latent variables~\cite{ma2019learning, tran2022aligning}.
In~\cite{ma2019learning}, the authors propose MacridVAE to disentangle user representations from macro- and micro-perspectives. The macro perspective may refer to the high-level concepts related to user intention, while the micro view may reflect the low-level of an item.
ADDVAE~\cite{tran2022aligning} extends MacridVAE by incorporating content information into their model.
In cross-modal retrieval (\eg image-text retrieval), MD-VAE~\cite{tian2022multimodal} is developed to disentangle the original {numeric} representations of each modality into modality-invariant and modality-specific features.
The inference models in the above work usually harness simple matrix factorization (MF) or multilayer perceptron (MLPs) to encode the rating or content information.
To encode high-order interactions in the user-item graph, MVGAE~\cite{yi2021multi} leverages the expressive power of VGAE~\cite{kipf2016variational} to tackle the data sparsity and uncertainty problem.
MVGAE also lacks interpretability as it is built on numerical embeddings from MLPs.
In this paper, we propose a graph variational auto-encoder (\dg{}) to exploit the high-order interactions in the user-item graph via a simplified residual GCN. 
Particularly, we design our model on the textual content to enhance its interpretability.

\section{Methodology}
\begin{figure*}
	\centering
	\includegraphics[width=0.92\textwidth, trim={50 10 50 5},clip]{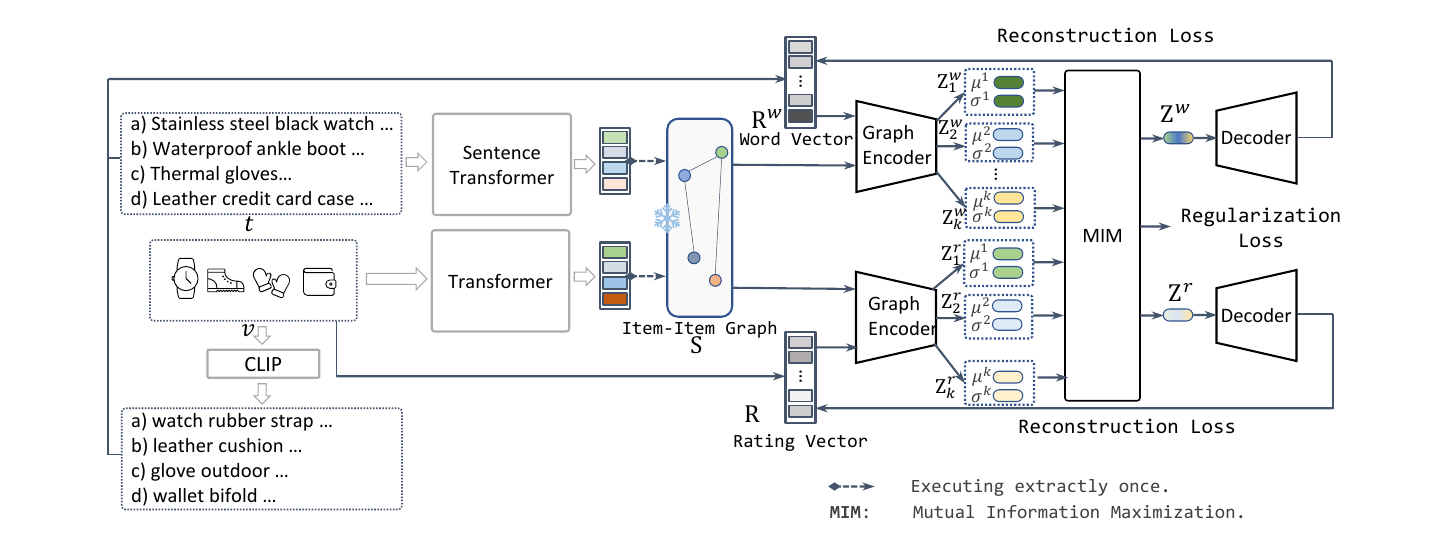} 
	\caption{{The framework of \dg{}, which fully utilizes multimodal information to construct the word vector and the item-item graph. \dg{} learns its model parameters by reconstructing both the word vector and the rating vector of a user. This figure is best viewed in color.}}
	\label{fig:framework}
\end{figure*}

Fig.~\ref{fig:framework} {illustrates} the overall architecture of \dg{}.
The inputs to \dg{} {consist of} user-item interactions and multimodal information.
To fully exploit multimodal information, we {employ} pre-trained multimodal models to transform information {from} other modalities into text. 
{Additionally}, we project raw modality-specific information into latent numerical representations.
{Building upon these} latent numerical representations, we construct an item-item graph and freeze it for performing graph convolutions. 

\subsection{Multimodal Information Preprocessing}
\label{sec:preprocess}

\textbf{Text-centered modality alignment.} 
In this paper, we adhere to established methodologies~\cite{liu2019user, zhang2021mining} by considering only visual and textual modalities, denoted by $\mathcal{M} = \{v, t\}$.
However, the proposed model can incorporate information from other modalities by {leveraging} pre-trained multimodal models~\cite{schneider2019wav2vec, radford2021learning}. 
With visual information, we use CLIP~\cite{radford2021learning} to align an image with {the} top-$k$ most relevant words describing this image.
Here, we denote the number of most relevant words extracted from images as a hyperparameter $Top_v$.
Finally, we obtain an item's multimodal textual representation by concatenating texts from all modalities.
As a result, we can further construct user textual preference by concatenating the textual representations of their interacted items.

\textbf{Multimodal numerical representations.}
Apart from the above alignment, we {utilize} a set of pre-trained models to transform modality-specific information into embeddings.
These embeddings {have proven} to be effective in multimodal recommendation~\cite{zhang2021mining}.
For the processing of images, we {employ a} pre-trained CNN to extract 4,096-dimensional visual embeddings from the image data~\cite{he2016ups}. 
In the processing of textual data, we employ sentence-transformers~\cite{reimers2019sentence} to extract a 384-dimensional textual embedding from the concatenation of the title, descriptions, categories, and brand of each item. 
We denote the numerical embedding from modality $m$ of an item $i$  as $\bm{x}_i^m$. We summarize the key notations used in our model with Table~\ref{tab:notation_table}.

\textbf{The construction of the homogeneous item graph.}
The aforementioned numerical embeddings are employed in the construction of a homogeneous item-item graph, which facilitates the exploitation of high-order semantic relationships between items. 
Specifically, we first use cosine similarity to compute the resemblance among
items within a specific modality $m$:
\begin{equation}
	\bm{S}_{i,j}^m = \frac{{(\bm{x}_i^m)}^\top \bm{x}_j^m}{ \| \bm{x}_i^m \| \| \bm{x}_j^m \|},
	\label{eq:sim}
\end{equation}
where $\bm{S}^m_{i,j}$ represents the element located in the $i$-th row and $j$-th column of the similarity matrix $\bm{S}^m \in \mathbb{R}^{N \times N}$. $N$ is the number of items.

The resultant similarity matrix $\bm{S}^m$ is dense and computationally inefficient. To address, we employ a $k$-nearest neighbor ($k$NN) strategy to trim unnecessary information and transform the matrix into a sparse representation. This process streamlines the computation and can be formally expressed as follows:
\begin{equation}
	\widehat{\bm{S}}^m_{i,j}=\begin{cases}
		1, \enspace & {\bm{S}}^m_{i,j} \in \operatorname{top-}k({\bm{S}}^m_{i}), \\
		0, \enspace & \text{otherwise}.
	\end{cases}
	\label{eq:topk}
\end{equation}
It is worth noting that the matrix $\widehat{\bm{S}}^m$ differs from the weighted similarity matrix {used} in LATTICE, which employs affinity values between items as its elements. In contrast, $\widehat{\bm{S}}^m$ represents the relationships between items using binary values, with 1 {indicates} the presence of a latent connection between two items. For each item $i$, only the connection relations of its top-$k$ most similar edges are retained. 

The discretized adjacency matrix $\widehat{\bm{S}}^m$ is then normalized as $\widetilde{\bm{S}}^m = ({\bm{D}^m})^{-\frac{1}{2}} \widehat{\bm{S}}^m ({\bm{D}^m})^{-\frac{1}{2}}$, where $\bm{D}^m$ is the diagonal degree matrix of $\widehat{\bm{S}}^m$ with dimensions $N \times N$. The diagonal elements of $\bm{D}^m$ are defined as $\bm{D}_{i,i}^m = \sum_{j}\widehat{\bm{S}}^m_{i,j}$.
We further utilize the resultant modality-aware adjacency matrices and construct the latent homogeneous item graph by aggregating the structures from each modality. 

\begin{equation}
	{\bm{S}} = \sum_{m \in \mathcal{M}} \alpha_{m} \widetilde{\bm{S}}^m,
\end{equation}
where \(\alpha_m\) denotes the importance score assigned to modality $m$, {and} $\mathcal{M}$ represents the set of all modalities under consideration. 
These parameters are used to weight the contributions of different modalities in the construction of the homogeneous item graph, {thereby enabling the} efficient representation of complex relationships between items in a dataset. 
{For simplicity, when considering two modalities}, we let \(\alpha_t = 1 - \alpha_v\).

In contrast to previous work, which dynamically updated the latent item-item graph~\cite{zhang2021mining}, we {choose} to fix it in order to achieve a significant improvement in the efficiency of \dg{}. 
{Freezing $\bm{S}$ eliminates the computational and memory costs that are quadratic} in the number of nodes during model training, resulting in a more efficient model.

\subsection{Disentangled Graph Variational Auto-Encoder}
On top of the constructed frozen item-item graph, \dg{} performs graph convolutional operators to reconstruct {the} users' rating matrix and their textual preferences.

\textbf{Definitions.}
Given a set of users $\mathcal{U}$, we denote the number of users as $M = |\mathcal{U}|$. 
The user-item interaction matrix {is} denoted as $\bm{R} \in \mathbb{R}^{M \times N}$, with each row {representing} a user's rating vector $\bm{r}_u \in \mathbb{R}^{1\times N}$.
We represent the user preference matrix constructed in multimodal preprocessing (section~\ref{sec:preprocess}) as $\bm{R}^w \in \mathbb{R}^{M \times W}$, where $W$ is the number of words.
Each element $r^w_{u,i} \in \bm{R}^w$ denotes user $u$'s preference {for} word $i$.
We denote the number of disentangled latent prototypes as $K$.
{The prototypes here may represent categories or brands of items.}

\textbf{Graph encoder.}
\dg{} encodes user interactions and item contents using GCNs across each disentangled prototype. 
The graph encoder in \dg{} takes either a user-item interaction matrix $\bm{R}$ or a user preference matrix $\bm{R}^w$ as input, and generates a list of latent variables $\bm{Z^r}=[\bm{Z}_1^r, \bm{Z}_2^r, \cdots, \bm{Z}_K^r]$ as output, where $\bm{Z}_k^r \in \mathbb{R}^{M \times d}$ and $d$ is the dimensional size. 
Here, we elaborate {on the process of \dg{} in} reconstructing the user-item rating matrix $\bm{R}$. The reconstruction of the user preference matrix $\bm{R}^w$ follows the same processes.
In $\bm{R}$, each row $\bm{r}_u \in \mathbb{R}^{1\times N}$ denotes user $u$'s rating vector.
The $k$-th component $\bm{z}_u^k \in \mathbb{R}^d$ represents user $u$'s preference on latent prototype $k$.
\begin{equation}
	q(\bm{Z}_k \mid \bm{R}) = \prod_{u=1}^M q(\bm{z}_u^k \mid \bm{R}),
\end{equation}
where $\bm{z}_u^k \in \bm{Z}_k^r$ is sampled from {a} multivariate Gaussian distribution:
\begin{equation}
	q(\bm{z}_u^k \mid \bm{R}) = \mathcal{N} \bigg(\bm{z}_u^k \bigm| \bm{\mu}_u^k, \text{diag}\left({(\bm{\sigma}_u^k)}^2 \right) \bigg).
	\label{eq:z_dis}
\end{equation}
Here, the mean $\bm{\mu}_u^k$ and the standard deviation $\bm{\sigma}_u^k$ are parameterized by a sequence model comprising a simplified residual GCN (Res-GCN) and a single layer MLP:
\begin{align}
	\label{eq:resgcn}
	& \bm{e}_u^k = \bm{r}_u \odot \bm{c}_{:,k}, \quad \bm{e}_u^k = \text{Res-GCN} \left(\frac{\bm{e}_u^{k}}{\left\|\bm{e}_u^{k}\right\|_2} \right),\\
	& \left(\bm{a}_u^k, \bm{b}_u^k\right) = \text{MLP}(\bm{e}_u^k), \\
	& \bm{\mu}_u^{k}=\frac{\bm{a}_u^{k}}{\left\|\bm{a}_u^{k}\right\|_2}, \quad \bm{\sigma}_u^{k} \leftarrow \sigma_0 \cdot \exp \left(-\frac{1}{2} \bm{b}_u^{k}\right),
\end{align}
where $\bm{e}_u^k \in \mathbb{R}^{1\times N}, \bm{a}_u^k \in \mathbb{R}^d, \text{and } \bm{b}_u^k \in \mathbb{R}^d$ are latent intermediate representations generated by Res-GCN and MLP.
{The term} $\bm{r}_u \odot \bm{c}_{:,k}$ {represents} the Hadamard product between the two vectors.
$\sigma_0$ is a hyperparameter.
$\bm{c}_{:,k} \in \mathbb{R}^{1 \times N}$ is the $k$-th column of the prototype matrix $\bm{C} \in \mathbb{R}^{N \times K}$.
Each row $\bm{c}_i \in \mathbb{R}^K$ in $\bm{C}$ represents the probabilities for this item belong to each disentangled prototype.
$\bm{c}_i$ is drawn from a categorical distribution:
\begin{align}
	& \bm{c}_i \sim \operatorname{CATEGORICAL} \left(\operatorname{SOFTMAX}([\rho_{i, 1}, \rho_{i, 2}, \ldots, \rho_{i, K}]) \right), \\
	\label{eq:rho_eq}
	& \rho_{i, k}= \bm{h}_i \bm{m}_k^\top / \tau,
\end{align}
where $\bm{m}_k \in \mathbb{R}^{1 \times d}$ is the representation of $k$-th prototype and $\bm{h}_i \in \mathbb{R}^{1 \times d}$ is the latent representation of an item (or word in $\bm{R}^w$). $\tau$ is a hyperparameter.

Given the frozen item-item graph $\bm{S}$, our simplified residual GCN (Res-GCN) in Eq.~\eqref{eq:resgcn} propagates graph convolutions for a user preference under category $k$ as:
\begin{equation}
	\bm{e}^k_{l} = \bm{e}^k_{l-1} {\bm{S}},
\end{equation}
where $\bm{e}^{k}_{l}$ is the $l$-th layer representation of a user intermediate preference.
When stacking $L$ layers, Res-GCN readouts the representation of an item by sum up its initial representation with its $L$-th hidden representation:
\begin{equation}
	\bm{e}^{k} = \bm{e}^k_0 + \bm{e}^k_L.
\end{equation}

\textbf{Decoder.}
Our generative model predicts the rating of user $u$ with regard to prototype $k$ over all items based on the latent variables $\bm{Z^r}$ and the prototype matrix $\bm{C}$. For simplicity, we omit the matrix identifier and denote user $u$'s latent variable as $\bm{z}^k_u$.
\begin{equation}
	p(\bm{r}_u | \bm{z}_u, \bm{C}) = \sum_{k=1}^K \exp\left( \bm{c}_{:,k} \cdot (\bm{z}_u^k \bm{H}^\top) / \tau \right),
\end{equation}
where $\bm{H}$ denotes the latent representation matrix of items in Eq.~\eqref{eq:rho_eq}.

\textbf{Regularizing disentangled prototypes with mutual information maximization (MIM).}
With the user-item interactions matrix $\bm{R}$ and user-word matrix $\bm{R^w}$, we can obtain two set of disentangled representations, $\bm{Z^r} = [\bm{Z}_1^{r}, \bm{Z}_2^{r}, \cdots, \bm{Z}_K^{r}]$ and $\bm{Z}^w = [\bm{Z}^w_{1}, \bm{Z}^w_{2}, \cdots, \bm{Z}^w_{K}]$, via the above graph encoder, respectively.

Following previous work~\cite{tay2019compositional, tran2022aligning}, we use a Compositional De-Attention to fuse the information from the two sets.
Given the sets of disentangled representations, the Compositional De-Attention computes an attentive score matrix $\bm{A} \in \mathbb{R}^{K \times K}$.
By alternatively {using} $\bm{Z}^r$ and $\bm{Z}^w$ as the key and query values, we obtain two attentive matrices, $\bm{A}_k^{r|w}$ and $\bm{A}_k^{w|r}$, respectively.
We then aggregate their disentangled representations with regard to $k$-th prototype as $\bm{Z}_k^{r|w} = \sum_{j=1}^K \bm{A}_{k,j}^{r|w} \bm{Z}_k^r$ and $\bm{Z}_k^{w|r} = \sum_{j=1}^K \bm{A}_{k,j}^{w|r} \bm{Z}_k^w$, respectively.

As we are primarily interested in maximizing the information between $K$ latent prototypes, we define a Jensen-Shannon MI estimator~\cite{hjelm2018learning} to minimize the following objective:
\begin{align}
	\mathcal{L}_{MI} = \sum_{k=1}^K & \Biggl(\mathbb{E}_\mathcal{U}[\operatorname{sp}(- \bm{Z}_k^{r|w} \odot \bm{Z}_k^{w|r})] + \nonumber \\ & \sum_{j=1, j\neq k}^K \mathbb{E}_\mathcal{U}[\operatorname{sp}(\bm{Z}_k^{r|w} \odot \bm{Z}_j^{w|r})] \Biggr),
\end{align}
where $\text{sp}(x) = \operatorname{ln}(1+e^x)$ is the softplus function.
This objective function encourages representations from the same prototype {to be} close to each other, while {repelling} representations from different prototypes.

\textbf{Optimization.}
We optimize the variational lower bound $\mathcal{L}_{LB}$ with {respect} to the parameters of Res-GCN and MLP by reconstructing the interaction matrix $\bm{R}$ (or $\bm{R}^w$):
\begin{align}
	\mathcal{L}_{LB} = \mathbb{E}_{p(\bm{C})} \bigr[& \mathbb{E}_{q(\bm{z}_u \mid \bm{R})}\left[\ln \left(p(\bm{r}_u \mid \bm{z}_u, \bm{C})\right)\right]-  \nonumber \\
	& \beta \cdot KL\left(q(\bm{z}_u \mid \bm{R}) \| p(\bm{z}_u)\right) \bigr],
\end{align}
where $q(\bm{z}_u \mid \bm{R})=\prod_{k=1}^K q(\bm{z}^k_u \mid \bm{R})$ and $\beta$ is a regularization term.
$\text{KL}[q(\cdot) \| p(\cdot)]$ is the Kullback-Leibler divergence between $q(\cdot)$ and $p(\cdot)$.
To optimize the approximate distribution in Eq.~\eqref{eq:z_dis}, we {utilize} the reparameterization trick~\cite{kingma2013auto} for training.
\begin{equation}
	\bm{z}(\bm{\mu}, \bm{\sigma}) = \bm{\mu} + \epsilon \odot \bm{\sigma},
\end{equation}
where $\epsilon \in \mathcal{N}(0, \bm{I})$.
The final learning objective is defined as:
\begin{equation}
	\mathcal{L} = \mathcal{L}_{LB} + \lambda \mathcal{L}_{MI}
\end{equation}
where $\lambda$ is a trade-off between the two {types} of losses.

\begin{table}[h]
	\centering
	\def\arraystretch{0.9}
	\setlength\tabcolsep{1.3pt} 
	\caption{{Notations used across the whole paper.}}
	\label{tab:notation_table}
	\begin{tabular}{ll}
		\hline
		\textbf{Symbols} & \textbf{Descriptions} \\
		\hline
		$\mathcal{M}$ & Set of item modalities, \ie visual and textual set {$v, t$} \\
		$\bm{x}_i^m$ & Numerical embedding of an item $i$ from modality $m$ \\
		$M, N$ & Number of of users, items \\
		$\bm{S}$ & The frozen item-item graph in matrix form \\
		$\bm{R}, \bm{R}^w$ & User-item interaction and user-word preference matrices \\
		$\bm{Z}^r, \bm{Z}^w$ & Disentangled latent representations on ratings and words \\
		$\bm{h}_i \in \bm{H}$& Learnable representation of an item $i$\\
		$\bm{m}_k$ & Learnable representation of a prototype $k$\\
		$\bm{c}_i \in \bm{C}$ & Categorical distribution of an item $i$ over prototypes\\
		$\bm{e}_u^k, \bm{a}_u^k, \bm{b}_u^k$ & A set of intermediate latent variables\\
		\hline
	\end{tabular}
\end{table}

\section{Experiments}
\label{sec:exp}
\subsection{Datasets}
We select three per-category datasets of the Amazon review dataset to evaluate our model as well as the baselines. Namely, (a) \emph{Baby}, (b) \emph{Sports and Outdoors} (denoted as \emph{Sports}), and (c) \emph{Clothing, Shoes and Jewelry} (denoted as \emph{Clothing}). The dataset of Aamzon review {has been} widely used in previous studies~\cite{zhang2021mining, zhou2023bootstrap}.
The raw data from each dataset undergoes a pre-processing procedure, which leverages a 5-core setting for both items and users. The results of this filtering process are presented in Table~\ref{tab:datasets}, providing a comprehensive overview of the data utilized in our analysis.

\begin{table}[bpt]
	\centering
	\def\arraystretch{1.25}
	\caption{Statistics of the experimental datasets.}
	\label{tab:datasets}
	\begin{tabular}{l r r r r}
		\hline
		\textbf{Dataset} & \textbf{\# Users} & \textbf{\# Items} & \textbf{\# Interactions} & \textbf{Sparsity} \\
		\hline
		Baby & 19,445 & 7,050 & 160,792 & 99.88\% \\
		Sports & 35,598 & 18,357 & 296,337 & 99.95\%\\
		Clothing & 39,387 & 23,033 & 278,677 & 99.97\%\\
		\hline
	\end{tabular}
\end{table}

\subsection{Evaluation Protocols}
For a fair comparison, we follow the same evaluation setting as~\cite{zhang2021mining, zhou2023bootstrap} with a random data splitting 8:1:1 on the interaction history of each user for training, validation, and testing.
We also adopt the widely use Recall@$K$ and NDCG@$K$ to evaluate the top-$K$ recommendation performance of our method as well as the baselines.
We empirically set $K$ to 10 and 20, {as in previous works}, and present the average metrics computed across all users in the test set.
We employ abbreviations to concisely report the average metrics of Recall@$K$ and NDCG@$K$ for all users in the test set, denoted as R@$K$ and N@$K$, respectively.

Formally, we define the metric of Recall@$K$ as:
\begin{equation*}
	\text{Recall}@K = \frac{1}{|\mathcal{U}^T|}\sum_{u\in \mathcal{U}^T }\frac{\sum^K_{i=1} \mathcal{I}[I^r_u(i) \in I^t_u]}{|I^t_u|},
\end{equation*}
where $\mathcal{U}^T$ represents the set of users encompassed within the test data, while $I^r_u(i)$ denotes the $i$-th item recommended for user $u$. The indicator function $\mathcal{I}[\cdot]$ is employed to count the number of recommended items that fall within the set of items $I^t_u$, which represents the items that have interactions with user $u$ in the testing data. 

We define NDCG@$K$ is defined as follows.
\begin{align*}
	& \text{NDCG}@K = \frac{1}{|\mathcal{U}^T|}\sum_{u\in \mathcal{U}^T }\frac{\text{DCG}@K(u)}{\text{IDCG}@K(u)}, \nonumber\\
	& \text{DCG}@K(u) = \sum^K_{i=1} \frac{2^{\mathcal{I}[I^r_u(i) \in I^t_u]}-1}{\log(i+1)},
\end{align*}
where $\text{IDCG}@K(u)$ denotes the ideal ranking scenario in which items that have interacted with user $u$ are positioned at the top. 

In addition to the aforementioned evaluations, we further assess the proposed model in the context of cold-start scenarios. Specifically, we adopt the widely studied item cold-start setting outlined in~\cite{yi2021multi}. In this setting, we randomly sample 20\% of items from the entire set and retain only two ratings per item within the training set. These items are then further divided into two equal subsets (10\% each), with their remaining interactions used for validation and testing. We further extend this analysis to the challenging zero-shot cold-start scenario, where no interactions are retained for the selected items in the training set.

\subsection{Comparative Baselines}
To demonstrate the effectiveness of \dg{}, we compare it with the following two categories of state-of-the-art recommendation methods. 

\textbf{General CF models.}
This category of models only utilize the user-item interactions for recommendation.
\begin{itemize}
	\item \textbf{Multi-VAE}~\cite{liang2018variational} improves the expressive power of linear factor models by extending VAEs into CF paradigms. Particularly, the model adopts a multinomial likelihood	function parameterized by neural network to model user-item interactions.
	\item \textbf{MacridVAE}~\cite{ma2019learning} is a generative model based on $\beta$-VAE that decouples the users' latent factors behind their decision marking processes.
	\item \textbf{DGCF}~\cite{wang2020disentangled} devises a graph disentangling module to refine the intent-aware graphs for disentangling factorial representations.
	\item \textbf{LightGCN}~\cite{he2020lightgcn} recommends items by simplifying the vanilla GCN with the removal of non-linear activation and feature transformation layers.
\end{itemize}

\textbf{Multimodal models.}
The models in this category either leverage numerical embeddings generated with pre-trained models or use word vectors represented by tf-idf values~\cite{li2017collaborative, zhu2022mutually}.
\begin{itemize}
	\item \textbf{VBPR}~\cite{he2016vbpr} extends the classic MF framework by incorporating visual features in BPR loss. Following previous work~\cite{zhang2021mining, wang2021dualgnn}, the multimodal features of an item are concatenated to form its visual feature, which is then utilized for the purpose of user preference learning.
	\item \textbf{MMGCN}~\cite{wei2019mmgcn} first performs graph convolutions within each modality of items, it then fuses the representations from all modalities for final prediction.
	\item \textbf{GRCN}~\cite{wei2020graph} learns a refined graph based on the representations of users and items. On top of the refined graph, it performs graph convolutions to obtain the representations of users and items.
	\item \textbf{DualGNN}~\cite{wang2021dualgnn} builds an auxiliary user-user correlation graph to augment the representations of users in GCNs.
	\item \textbf{LATTICE}~\cite{zhang2021mining} exploits graph structure learning to explicitly learn the latent semantic relations between items based on their multimodal features. The model performs graph convolutions on both user-item bipartite graph and the built item-item graph for recommendation. 
	\item \textbf{MVGAE}~\cite{yi2021multi} employs VGAE~\cite{kipf2016variational} to obtain modality-specific embeddings of nodes and fuses the embeddings for recommendation. The vanilla version, leveraging the product-of-experts principle, actually led to performance degradation, so we evaluate a variant of MVGAE with the removal of its PoE component, which we dubbed as \textbf{MVGAE{\footnotesize (w/o POE)}}. This variant is closely related to VGAE; however, MVGAE{\footnotesize (w/o POE)} performs graph convolutions on a bipartite user-item graph.
	\item \textbf{SLMRec}~\cite{tao2022self} proposes the implementation of three data augmentation techniques in self-supervised learning to extract the multimodal patterns present in data for the purpose of multimedia recommendation.
        \item \textbf{BM3}~\cite{zhou2023bootstrap} introduces a novel contrastive learning approach to bootstrap the representation of users and items for multimodal recommendation, eliminating the need for negative samples.
        \item \textbf{DMRL}~\cite{liu2022disentangled} disentangles representations within individual modalities and employs an attention module to determine users’ preferences for these representations, thereby enhancing recommendation accuracy.
\end{itemize}

\begin{table*}[bt]
	\centering	
	\def\arraystretch{0.96}
	\caption{The overall performance of various recommendation methods is compared in terms of Recall and NDCG metrics. The global best results for each dataset and metric are denoted in \textbf{boldface}, while the second-best results are \underline{underlined}. The improvement percentage, denoted as \emph{improv.}, is calculated as the ratio of performance increment from the best baseline to \dg{} for each dataset and metric. To ensure the stability of our method, experiments were conducted across 5 different seeds, and the improvements were found to be statistically significant at a level of $p < $ 0.01 using a paired $t$-test. *}
	\begin{tabular}{l cccc cccc cccc}
		\hline
		\textbf{Dataset} & \multicolumn{4}{c}{\textbf{Baby}} & \multicolumn{4}{c}{\textbf{Sports}} & \multicolumn{4}{c}{\textbf{Clothing}} \\
		\hline
		\textbf{Metric} & \textbf{R@10} & \textbf{R@20} & \textbf{N@10} & \textbf{N@20} & \textbf{R@10} & \textbf{R@20} & \textbf{N@10} & \textbf{N@20} & \textbf{R@10} & \textbf{R@20} & \textbf{N@10} & \textbf{N@20} \\
		\hline
		Multi-VAE & 0.0353 & 0.0579 & 0.0189 & 0.0248 & 0.0420 & 0.0620 & 0.0230 & 0.0282 & 0.0232 & 0.0358 & 0.0123 & 0.0156 \\
		MacridVAE & 0.0463 & 0.0703 & 0.0264 & 0.0327 & 0.0558 & 0.0811 & 0.0325 & 0.0390 & 0.0352 & 0.0518 & 0.0195 & 0.0237 \\
		DGCF & 0.0441 & 0.0709 & 0.0239 & 0.0308 & 0.0515 & 0.0774 & 0.0285 & 0.0352 & 0.0311 & 0.0474 & 0.0168 & 0.0210 \\
		LightGCN & 0.0479 & 0.0754 & 0.0257 & 0.0328 & 0.0569 & 0.0864 & 0.0311 & 0.0387 & 0.0361 & 0.0544 & 0.0197 & 0.0243 \\
		\hline
		VBPR & 0.0423 & 0.0663 & 0.0223 & 0.0284 & 0.0558 & 0.0856 & 0.0307 & 0.0384 & 0.0281 & 0.0415 & 0.0158 & 0.0192 \\
		MMGCN & 0.0421 & 0.0660 & 0.0220 & 0.0282 & 0.0401 & 0.0636 & 0.0209 & 0.0270 & 0.0227 & 0.0361 & 0.0120 & 0.0154 \\
		GRCN & 0.0532 & 0.0824 & 0.0282 & 0.0358 & 0.0599 & 0.0919 & 0.0330 & 0.0413 & 0.0421 & 0.0657 & 0.0224 & 0.0284 \\
		DualGNN & 0.0513 & 0.0803 & 0.0278 & 0.0352 & 0.0588 & 0.0899 & 0.0324 & 0.0404 & 0.0452 & 0.0675 & 0.0242 & 0.0298 \\
		LATTICE & {0.0547} & {0.0850} & {0.0292} & {0.0370} & 0.0620 & 0.0953 & {0.0335} & {0.0421} & {0.0492} & {0.0733} & {0.0268} & {0.0330} \\
		MVGAE{\footnotesize (w/o POE)} & 0.0307 & 0.0520 & 0.0151 & 0.0206 & 0.0246 & 0.0446 & 0.0116 & 0.0167 & 0.0207 & 0.0337 & 0.0104 & 0.0137 \\
		SLMRec & 0.0521 & 0.0772 & 0.0289 & 0.0354 & {0.0663} & {0.0990} & {0.0365} & {0.0450} & 0.0442 & 0.0659 & 0.0241 & 0.0296 \\
		BM3 & \underline{0.0564} & \underline{0.0883} & {0.0301} & {0.0383} & 0.0656 & 0.0980 & 0.0355 & 0.0438 & 0.0422 & 0.0621 & 0.0231 & 0.0281 \\
		DMRL & 0.0543 & 0.0847 & \underline{0.0322} & \underline{0.0405} & \underline{0.0672} & \underline{0.1008} & \underline{0.0393} & \underline{0.0484} & \underline{0.0549} & \underline{0.0791} & \underline{0.0311} & \underline{0.0373} \\	
		\hline
		\dg{} & \textbf{0.0636} & \textbf{0.1009} & \textbf{0.0340} & \textbf{0.0436} & \textbf{0.0753} & \textbf{0.1127} & \textbf{0.0410} & \textbf{0.0506} & \textbf{0.0619} & \textbf{0.0917} & \textbf{0.0336} & \textbf{0.0412} \\	
		\emph{improv.} & 12.77\% & 14.27\% & 5.59\% & 7.65\% & 12.05\% & 11.81\% & 4.33\% & 4.55\% & 12.75\% & 15.93\% & 8.04\% & 10.46\% \\	
		\hline
	\end{tabular}
	\label{tab:perform}	
\end{table*}

\subsection{Implementation and Hyperparameter Settings}
To ensure a fair comparison and align our work with established methodologies delineated in previous researches~\cite{he2020lightgcn, zhang2021mining}, we fixed the dimensionality of the embeddings for both users and items at 64 for all models. The initialization of the embedding parameters was performed utilizing the Xavier method~\cite{pmlr-v9-glorot10a}, while the Adam algorithm~\cite{kingma2015adam} was employed as the optimizer. Furthermore, we meticulously adjusted the parameters of each model in accordance with the specifications outlined in their respective publications.
The number of GCN layers in the item-item graph is fixed at 2 for reduction of the hyperparameter searching space in \dg{}.
We empirically set the number of words $Top_v$ fused from images at $5$, both the visual feature ratio $\alpha_v$ and $\beta$ at 0.1.
A grid search is conducted on the hyperparameters of \dg{} across all datasets to determine its optimal settings.
Specifically, the number of latent prototypes in \{3, 4, 5\} and $\tau = 0.1$.
The regularization term $\lambda$ in \{0.1, 0.2, 0.3, 0.4, 0.5\}.
For the purpose of convergence, we set the early stopping epochs at 20.
In accordance with the methodology outlined in~\cite{zhang2021mining}, the R@20 metric on the validation data is utilized as the training stopping indicator.
For multimodal baselines, we employ the implementation encapsulated within the MMRec library~\cite{zhou2023mmrec}.

\begin{figure*}[bpt]
	\centering
	\includegraphics[width=0.98\textwidth, trim={10 0 10 10},clip]{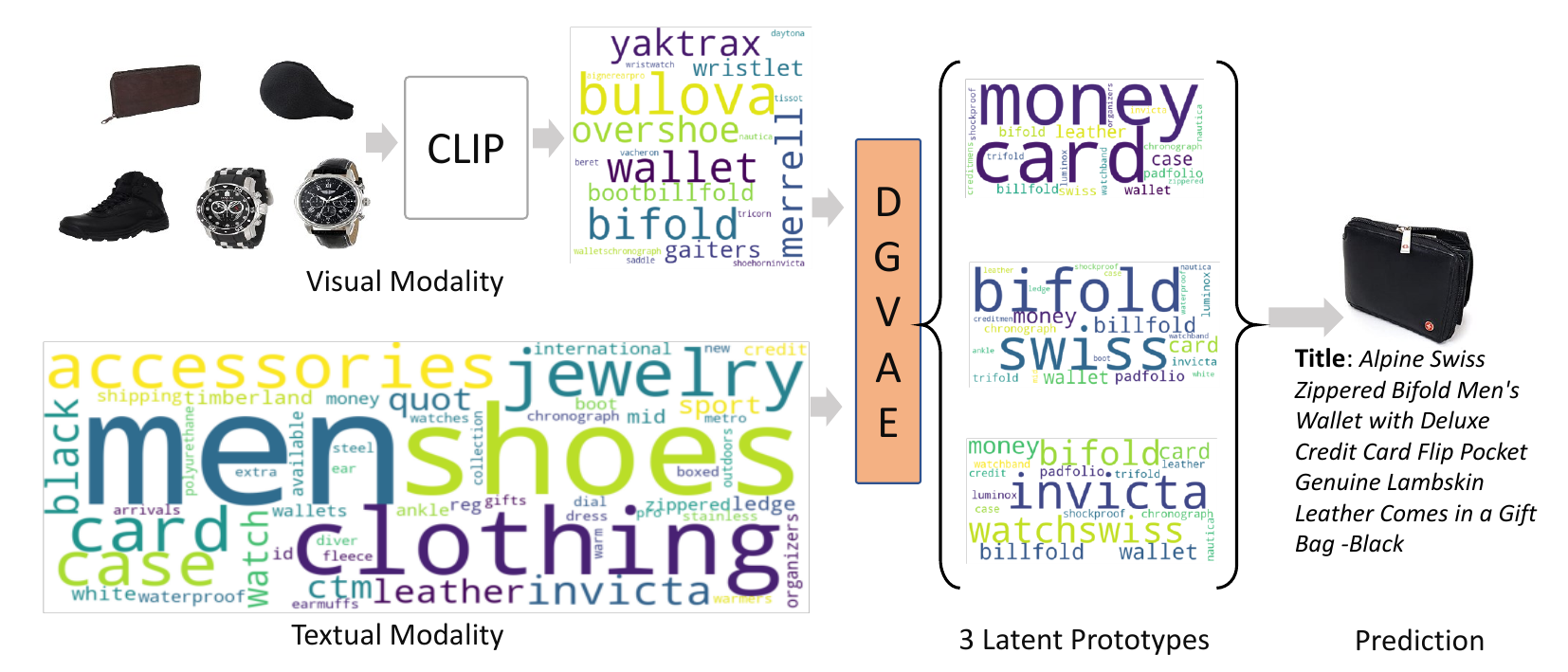} 
	\caption{An illustration of interpretability of \dg{}. The left part shows the interacted items by user ``A1E3O99XB3BN3W''. The middle part visualizes the learned latent prototypes of \dg{}. The right part presents the learned prototypes and the recommended item.}
	\label{fig:dgvae-pre}
\end{figure*}

\subsection{Performance Comparison}
\textbf{Overall performance.}
Table~\ref{tab:perform} shows the overall performance with regard to Recall and NDCG obtained by the baselines. 
\textit{First}, we observe that graph-based models, encompassing both general collaborative filtering and multimodal categories, exhibit competitive performance in recommendation accuracy.
Graph-based models are capable of exploiting the high-order interactions of users to alleviate the data sparsity.
As a result, our proposed \dg{} harnessing the item-item graphs can outperform the best baselines, BM3 and DMRL, across all evaluation metrics.
\textit{Second}, disentangled MacridVAE obtains better performance over Multi-VAE. The performance of MacridVAE is superior to the disentangled graph model DGCF, which perform graph convolutions on user-item interactions. Analogously, DMRL exhibits superior performance among all baselines across the majority of evaluation metrics. The results demonstrate the powerful expressiveness of disentangled VAEs in recommendation.
\textit{Third}, with the exception of MMGCN, models incorporating multimodal information demonstrate a significant performance improvement compared to those lacking such information. For MMGCN, we speculate the reason may result from their graph convolutional operators are merely performed on raw user-item graph. Whilst the line of graph-based multimodal models (\eg GRCN, DualGNN, LATTICE) either refine the raw graph or leverage auxiliary graphs of user-user or item-item for graph convolutions. Whilst other models (\eg SLMRec, BM3) injects contrastive signals into user-item interactions for effective learning.
DMRL~\cite{liu2022disentangled} effectively leverages multimodal information and disentangles the representations, resulting in superior performance in the majority of cases when compared to the baselines. 
\dg{} freezes the item-item graph for training also achieve competitive recommendation accuracy.
\textit{Fourth}, it is worth noting that the performance of MVGAE is unacceptably poor. 
As MVGAE is built upon MMGCN, we remove its PoE component and find MVGAE{\footnotesize (w/o POE)} can boost the performance of vanilla MVGAE by nearly 10$\times$.
To illustrate the performance degradation in MVGAE, we resort to the definition of PoE in~\cite{hinton2002training}.
\begin{equation}
	P(x | \{\theta_m\})=\frac{1}{\int \mathrm{d}x \prod_{m\in \mathcal{M}} f_m(x |\theta_m)} \prod_{m\in \mathcal{M}} f_m (x | \theta_m), 
\end{equation}
where $f(x)$ is an expert defined as a probabilistic model {that represents} the input space or modality in MVGAE. 
The definition of PoE requires the probabilities from each modality should {be} prominent to obtain a good result. In a multimodal context, MVGAE requires all multimodal features, calculated by experts, to be prominent {in order to} achieve an informative final item representation.
However, in practice, this requirement may be difficult to satisfy, potentially leading to a degradation in the recommendation quality of the vanilla MVGAE.
The removal of PoE in MVGAE makes it comparable to MMGCN in {terms} of magnitude, although it is still inferior to MMGCN.
\textit{Lastly}, our proposed model, \dg{}, {benefits} from both graph convolutions and disentangled VAEs, {achieving} the best recommendation accuracy across three datasets.
We {will} illustrate its interpretability in the following sections. 

\begin{figure}[tbp]
	\centering
	\includegraphics[width=\linewidth, trim={0 0 0 0},clip]{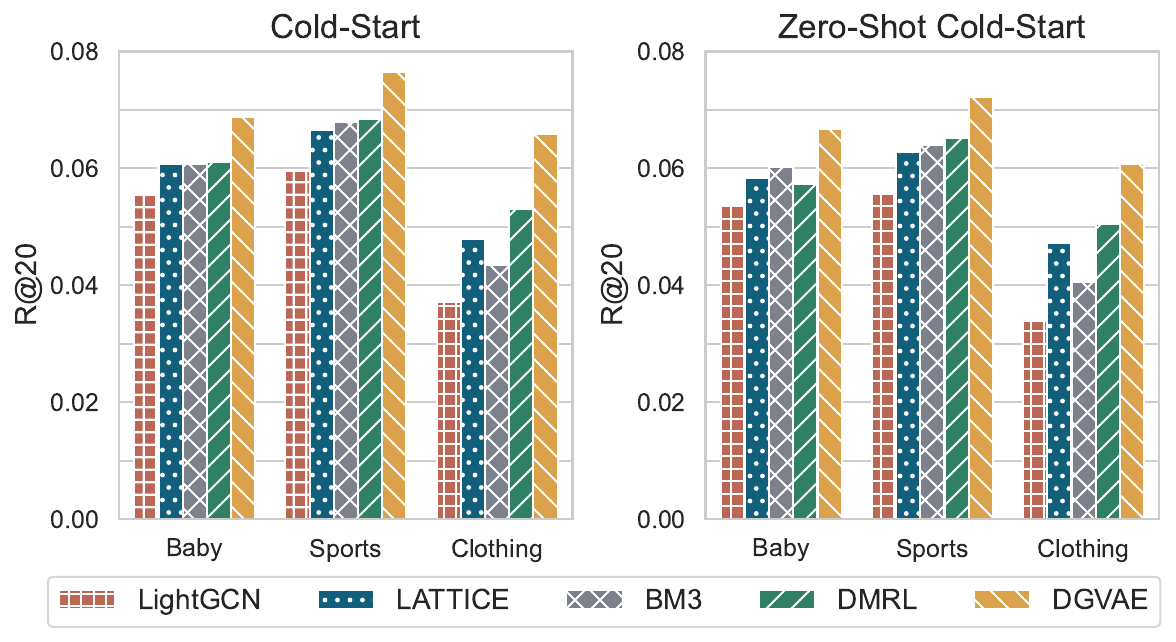} 
	\caption{Performance of \dg{} compared with various baselines under different cold-start settings.}
	\label{fig:cold}
\end{figure}

\textbf{Performance in cold-start settings.}
Fig.~\ref{fig:cold} delineates a comparative analysis of the performance (\ie R@20) for \dg{} and a selection of key baseline models under various cold-start scenarios. Intuitively, one would anticipate that multimodal models, given their ability to harness a comprehensive array of content information, would exhibit superior performance in recommendation tasks. This intuitive expectation is empirically validated in our observations, as depicted in Fig.~\ref{fig:cold}, where the general recommendation model, LightGCN, demonstrates a performance that is noticeably inferior when contrasted with the multimodal models.
Compared to the zero-shot cold-start scenario, models trained with even a few more ratings (\eg two per selected item) exhibited significant improvements in performance across all tested algorithms. This suggests that access to even minimal user interaction data facilitates more accurate inference of user preferences, leading to enhanced performance for recommendation models.
Furthermore, Fig.~\ref{fig:cold} reveals that our proposed model demonstrates exceptional efficacy in cold-start situations. This is primarily attributed to the capabilities of \dg{}, which adeptly captures the interrelationships among items through two key mechanisms: the exploitation of the rich content inherent in items via a high-order item-item graph, and the reconstruction of item textual semantics using a VAE. Remarkably, the results are obtained even in the absence of user-item interactions, underscoring the robustness of our model. In our future research, we aim to further enhance the resilience of \dg{} in the context of a code-start scenario, particularly within the framework of sequential recommendation systems~\cite{zhang2023multimodal, zhang2024id}.

\vspace{-10pt}

\subsection{Interpretability}
\label{sec:interp}
We analyze the interpretability of the proposed model and recommendation results by exemplifying a user instance with the ID of ``A1E3O99XB3BN3W'' {from} the Clothing dataset.
This user has 5 interactions in the training set and 1 interaction in the test set.
The detailed descriptions and cover images of the products this user interacted with, which are hosted on Amazon, are listed in Table~\ref{tab:product_details} in Appendix~\ref{append:product}.
Fig.~\ref{fig:dgvae-pre} shows the visual and textual information of these interacted products. 
With raw cover images, we use a pre-trained multimodal model CLIP~\cite{radford2021learning} to transform the images into words.
The left part of Fig.~\ref{fig:dgvae-pre} presents the word clouds of visual and textual modalities, where the size of the words is proportional to their frequency in the input text.
In the visual modality, we observe that the words ``bifold'' and ``wallet'' are frequently used to describe the cover images of the products. 
However, these words do not { appear frequently} in the {textual} modality. 
The right part of Fig.~\ref{fig:dgvae-pre} shows the products recommended by \dg{} and purchased by this user.

\textbf{Interpretability of recommendation results.}
For this user (\ie A1E3O99XB3BN3W), \dg{} recommends a wallet which shows in the right of Fig.~\ref{fig:dgvae-pre}. 
When we compare the recommended product with the products this user {has} interacted with, we can interpret the user's preference on this product (\ie B004M6UDC8).
Thanks to the multimodal pre-trained model, the title of the recommended product is highly related {to} the visual information extracted from the cover images of previously interacted products.
Previous multimodal models leveraging the numerical embeddings of products are difficult to understand the connections between recommended products and the interacted products.

\textbf{Model interpretability.}
We conduct a further investigation into why \dg{} predicts the product of ``B004M6UDC8'' for user ``A1E3O99XB3BN3W''.
We visualize the top predicted words under each latent prototype in the middle of Fig.~\ref{fig:dgvae-pre}.
The three latent prototypes in the figure can {be differentiated} from each other with the help of mutual information theory.
To be specific, the top prototype highlights some product related to money and {cards, but pays less attention to} a bifold wallet.
{Conversely}, the second prototype focuses on a bi-fold {item from a} Swiss brand.
The last prototype shows interest in topics related to Swiss watches.

Fig.~\ref{fig:dgvae-pre} also reveals that the latent prototypes may learn some counter-intuitive representations.
\textit{Firstly}, the frequently occurred words (\eg men, clothing, shoes) in the textual modality may not be captured by any latent prototype. 
The reason is that these words have commonly appeared in the Clothing dataset, thus, the importance of these words is understated with tf-idf.
\textit{Secondly}, we observe some words (\eg Swiss) do not exist in both the visual and textual modalities but {are attended} to the latent prototypes of the user's preference. 
We speculate the reason may result from the graph convolutional operators of \dg{}.
Because graph convolutions can learn a node's representation from its neighbors. Take the item of watch in Fig.~\ref{fig:dgvae-pre} for example, \dg{} may learn word ``Swiss'' from other similar watches that tagged with ``Swiss'' label.
Thus, \dg{} use the word ``Swiss'' to describe the watch purchased by user ``A1E3O99XB3BN3W''.
{Thanks to} the graph convolutional layer of \dg{}, even items with few interactions can learn rich semantic information from their neighbors.

\subsection{Ablation Study}
We carry out a series of ablation studies, which involved the process of ablating the item-item graph, the ablating of the mutual alignment between ratings and multimodal features in \dg{}, as well as the impact of multimodal features with regard to performance of \dg{}, etc.

\begin{figure}[tbp]
	\centering
	\includegraphics[width=\linewidth, trim={0 0 0 0},clip]{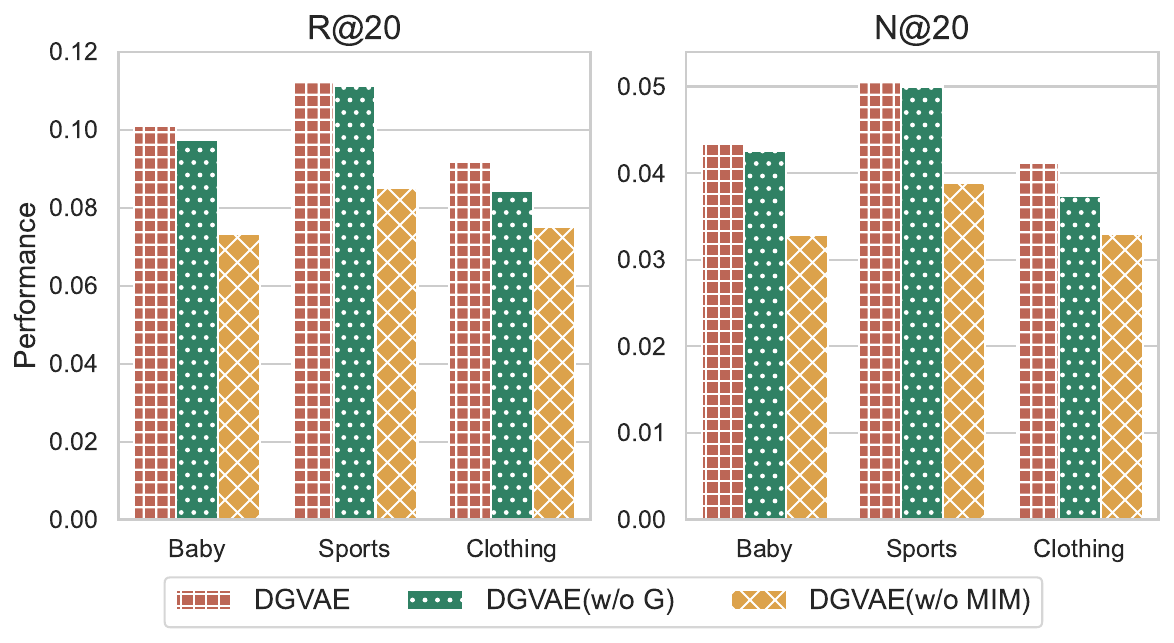} 
	\caption{Comparison of \dg{} with its variants.}
	\label{fig:plot-ab}
\end{figure}

\textbf{Components of \dg{}.}
Fig.~\ref{fig:plot-ab} shows the performance of \dg{} with the following variants:
\begin{itemize}
\item \textbf{\dg{}(w/o G)} is a variant of \dg{} without the item-item graph.
\item \textbf{\dg{}(w/o MIM)} is a variant of \dg{} without the regularization of mutual information maximization.
\end{itemize}
The result shows that the performance of \dg{}(w/o G) is slightly worse than its full version on Baby and Sports. 
Specifically, the R@20 for \dg{} improved over \dg{}(w/o G) are 3.5\% and 1.3\% on Baby and Sports.
However, we observe a significant performance boost on the largest dataset, Clothing.
To be specific, the improvements are 8.9\% on R@20 and 10.2\% on N@20.
We posit that the observed phenomenon is likely attributable to the effective utilization of latent multimodal relationships among items. These relationships are leveraged in the assembly of an item-item graph, which serves as a substantial supplement to user-item interactions. This approach demonstrates particular efficacy in the context of datasets characterized by a sparse quantity of user interactions.
Moreover, it is noteworthy that \dg{}(w/o MIM) exhibits the least effective performance, underscoring the critical role of mutual information maximization in the functionality of \dg{}.
However, \dg{}(w/o MIM) exhibits a marginal enhancement over the disentangled collaborative filtering method, MacridVAE, even in the absence of aligning ratings and multimodal information via mutual information maximization.

\begin{figure}[tbp]
	\centering
	\includegraphics[width=\linewidth, trim={0 0 0 0},clip]{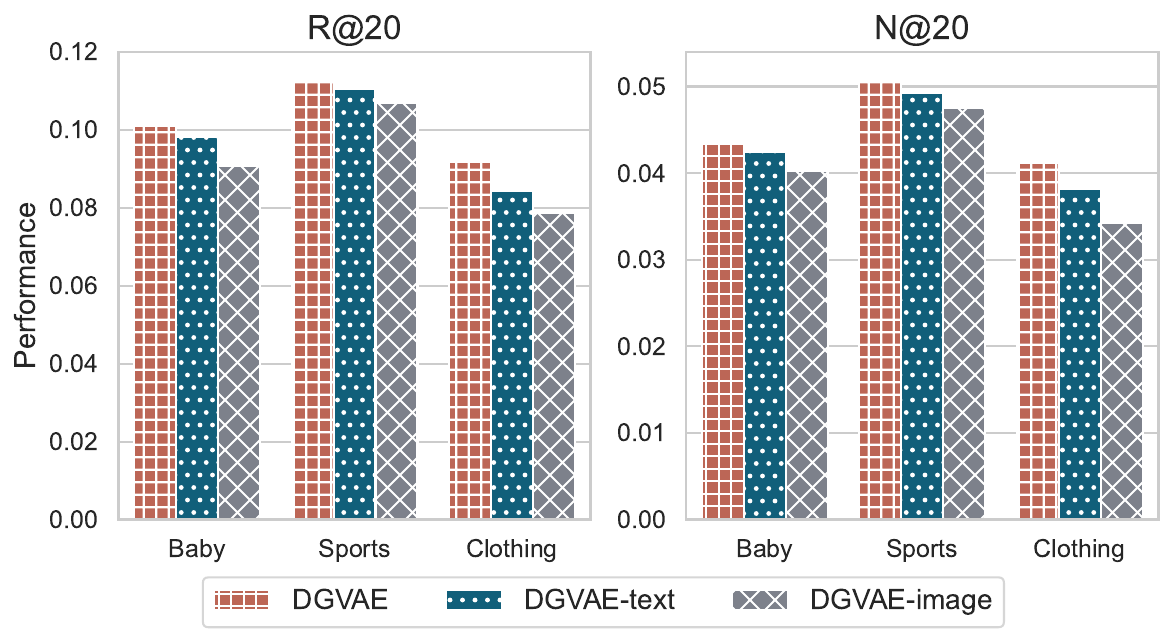} 
	\caption{Comparison of \dg{} under different uni-modal features.}
	\label{fig:plot-ab-mm}
\end{figure}

\textbf{Multimodal features in \dg{}.}
We further examine the impact of various uni-modal information on the performance of \dg{}. We have devised the following variants of \dg{}:
\begin{itemize}
\item \textbf{\dg{}-text} is a variant of \dg{} that only utilizes the \textbf{text} information associated with items.
\item \textbf{\dg{}-image} is a variant of \dg{} that only utilizes the \textbf{image} information associated with items.
\end{itemize}
The resulted Fig.~\ref{fig:plot-ab-mm} indicates that within the \dg{} framework, the text modality holds greater importance than the image modality. Moreover, when juxtaposed with the performance metrics in Table III of the main manuscript, it is observed that the \dg{} framework, when utilizing solely on single modality information, is on par or even surpasses the performance of the best baselines.

\textbf{Visual pre-trained model for image conversion.}
Rather than employing the CLIP~\cite{radford2021learning} for image-to-text conversion, we have incorporated BILP-2~\cite{li2023blip} into our proposed model, \dg{}. We then assess the impact of different image-to-text encoders on its performance, shown in the following Table~\ref{tab:clip_blip}.

\begin{table}[h]
\caption{Comparative analysis of \dg{} with various image-text encoders. The best results are in \textbf{boldface}.}
\def\arraystretch{.9}
\label{tab:clip_blip}
\begin{tabular}{llllll}
\hline
Dataset                   & image-text encoders & R@10            & R@20            & N@10            & N@20            \\ \hline
\multirow{2}{*}{Baby}     & CLIP                & 0.0636          & 0.1009          & 0.0340          & 0.0436          \\ \cline{2-6} 
                          & BLIP-2              & \textbf{0.0641} & \textbf{0.1011} & \textbf{0.0346} & \textbf{0.0441} \\ \hline
\multirow{2}{*}{Sports}   & CLIP                & 0.0753          & 0.1127          & 0.0410          & 0.0506          \\ \cline{2-6} 
                          & BLIP-2              & \textbf{0.0765} & \textbf{0.1146} & \textbf{0.0418} & \textbf{0.0516} \\ \hline
\multirow{2}{*}{Clothing} & CLIP                & 0.0619          & 0.0917          & 0.0336          & 0.0412          \\ \cline{2-6} 
                          & BLIP-2              & \textbf{0.0622} & 0.0910          & \textbf{0.0341} & \textbf{0.0414} \\ \hline
\end{tabular}
\end{table}

The results presented in the table suggest that the \dg{} model, when equipped with BLIP-2, exhibits a marginally superior performance compared to the CLIP model. This observation implies that an enhanced image-text encoder can potentially lead to an improved image representation, thereby augmenting the recommendation performance. However, as depicted in Fig.~\ref{fig:plot-ab-mm} of the revised manuscript, the performance of the \dg{} model utilizing image data is not comparable to that of text data. Consequently, the performance improvement attributed to the superior image representation of BLIP-2 is relatively limited, as evidenced in the aforementioned table.

\subsection{Hyperparameter Sensitivity Study}
In this section, we study how hyperparameters in \dg{} would affect its performance.
Specifically, we evaluate \dg{} with the number of latent prototypes in \{2, 3, 4, 5, 6\} and the number of words attending into textual information from images varying in \{0, 5, 10, 15, 20\}.

\begin{figure}[tbp]
	\centering
	\includegraphics[width=\linewidth, trim={10 10 10 10},clip]{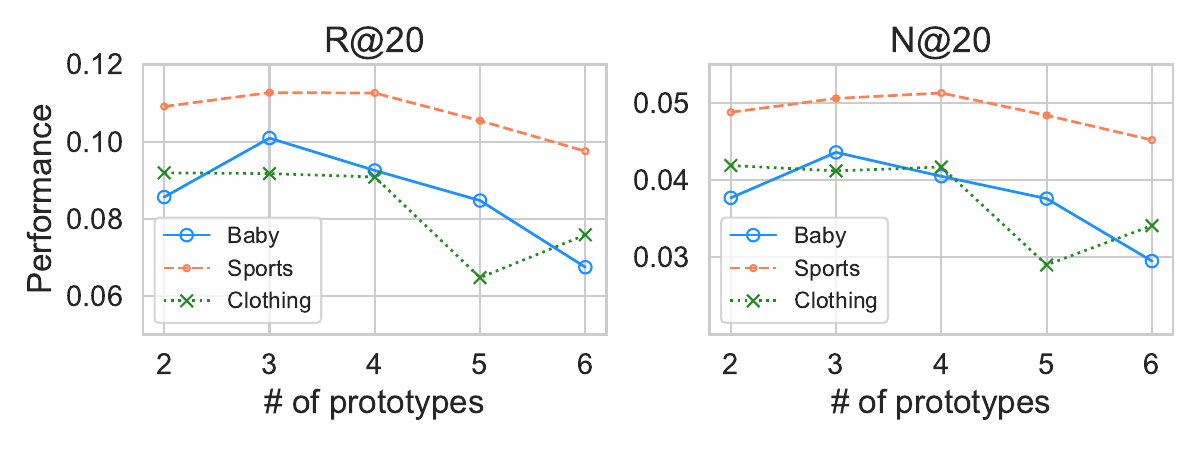} 
	\caption{Performance of \dg{} varies with the number of disentangled prototypes.}
	\label{fig:plot-ncl}
\end{figure}

\textbf{Number of prototypes.} 
We plot the performance of \dg{} with respect to R@20 and N@20 under different numbers of prototypes in Fig.~\ref{fig:plot-ncl}.
Across all datasets, we observe a performance degeneration when disentangling too many prototypes in \dg{}. 
On the contrary, too few prototypes equipped in \dg{} may not possible to decouple the disentangled user representations.
Therefore, the results in Fig.~\ref{fig:plot-ncl} suggest that 3 or 4 prototypes would be a relational choice for \dg{}.

\begin{figure}[tbp]
	\centering
	\includegraphics[width=\linewidth, trim={10 10 10 10},clip]{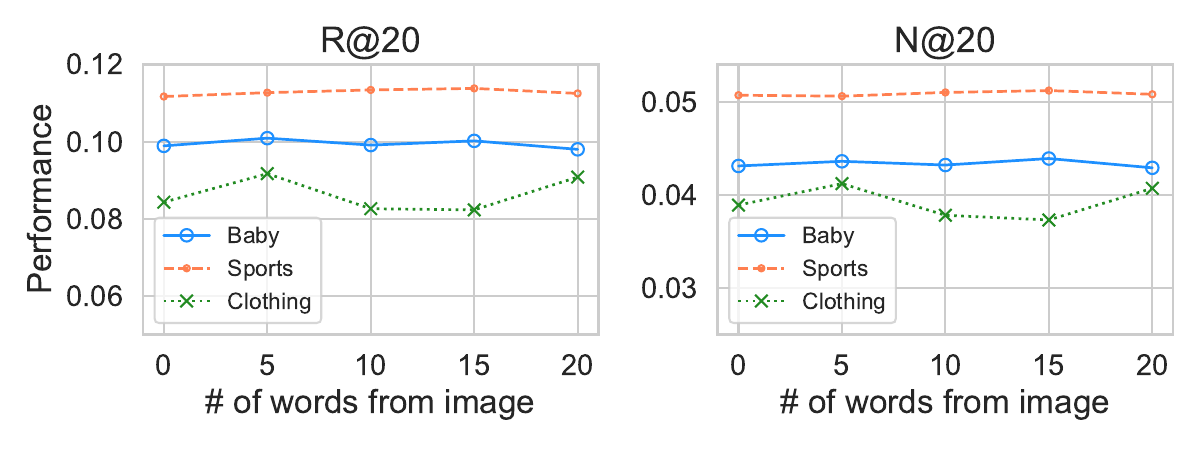} 
	\caption{Performance of \dg{} varies with the number of words extracted from images.}
	\label{fig:plot-textno}
\end{figure}

\textbf{Number of words extracted from images.} 
In \dg{}, we use a pre-trained multimodal model CLIP~\cite{radford2021learning} to extract the most related words from the cover of a product. 
\dg{} fuses the $Top_v$ related words ranked by CLIP into the textual descriptions of items.
We plot the performance of \dg{} varying with the number of words in Fig.~\ref{fig:plot-textno}.
It is worth noting that ``0'' value in $x$-axis means \dg{} does not use visual multimodal information for recommendation.
The figure reveals that the visual information generated from CLIP places a limited impact on the performance of \dg{}, especially on Baby and Sports.
The dataset of Clothing is more sensitive to the number of words expressed by images fused in \dg{}.
This observation is in line with the findings reported in~\cite{zhang2021mining}.

To quantify the contribution of visual modality, we calculate the averaged improvements of \dg{} incorporating visual information on R@20 and N@20 as 3.9\% and 2.3\%, respectively.
The improvement in Clothing is particularly significant, reaching 8.8\% on R@20.
The results match the intuition that visual features are informative for multimodal models with the vision-sensitive dataset.

\subsection{Discussions and Limitations}
\textbf{Failure recommendation case of \dg{}.}
Since no recommendation model can predict all users’ responses perfectly, we were intrigued by the cases where collaborative signals can accurately predict outcomes, yet multimodal recommenders (i.e., \dg{}) do not succeed.
Consequently, we conducted an analysis of the recommendation results produced by LightGCN~\cite{he2020lightgcn} and our proposed \dg{} on the Clothing dataset. 

For simplicity consideration, our analysis is focused on users who have only one instance in the test data of the Clothing dataset. This subset comprises 37,937 users out of a total of 39,387 users. Remarkably, within this group of 37,937 users, there exist 2019 users where \dg{} accurately predicts their preference, whereas LightGCN does not. And there are 586 users where LightGCN can accurately predicts their preference, whereas \dg{} does not.

Among these 586 users, we exemplified one representative user (``A3JJ2LAPAM96FJ") and his/her interacted items in training/test/recommended data of \dg{} in the subsequent Table~\ref{tab:fail_case}.
The table illustrates that the system, denoted as \dg{}, recommended a fishing shirt to the user, a recommendation derived from the user's previous interactions with swim board shorts and t-shirts. The presence of multimodal information may potentially influence the collaborative signal, causing it to deviate from its intended direction. Therefore, adapting the alignment of multimodal features to correspond with users' preferences remains a significant scientific challenge in multimodal recommendation.

\begin{table}[H]
\caption{Failure case of \dg{}.}
\label{tab:fail_case}
\def\arraystretch{.9}
\begin{tabular}{l|l|l}
\hline
\textbf{Items in Training} & \textbf{Item(s) in Test}   & \textbf{Top-recommendation} \\ \hline
\begin{tabular}[c]{@{}l@{}}ID: B00009ZM7Z\\ Title: merrell men's \\jungle moc slip-on shoe\end{tabular}                                    & \multirow{10}{*}{\begin{tabular}[c]{@{}l@{}}ID: B0018OM1TU\\ Title: levi's men's \\559 relaxed straight \\leg jean\end{tabular}} & \multirow{10}{*}{\begin{tabular}[c]{@{}l@{}}ID: B0002XSXWC\\ Title: columbia men's \\bonehead short sleeve \\fishing shirt \end{tabular}} \\ \cline{1-1}
                                \begin{tabular}[c]{@{}l@{}}ID: B00AN53FMW\\ Title: hurley men's \\one and only 22 inch \\supersuede boardshort \\swim board shots\end{tabular} &    &                                                                                                                        \\ \cline{1-1}
                                \begin{tabular}[c]{@{}l@{}}ID: B00BJJJA1G\\ Title: levi's men's \\grosevelt t-shirt, charcoal\\ heather, x-large, t-shirts\end{tabular}      &    &                                                                                                                        \\ \hline
\end{tabular}
\end{table}

\textbf{Discussion of popular bias in \dg{}.}
In \dg{}, text or words serve a dual purpose: they are utilized to construct the item-item graph and to align user ratings via word vectors. Hence, we discuss the influence of popularity bias among the textual words from two distinct viewpoints.
\begin{itemize}
\item \textbf{TF-IDF (Term Frequency–Inverse Document Frequency) text vectorization}. \dg{} employs word vectors characterized by Term Frequency-Inverse Document Frequency (TF-IDF) values. The TF-IDF scoring mechanism quantifies a word by multiplying the Term Frequency (TF) of the word by its Inverse Document Frequency (IDF)~\cite{zhu2022mutually}. In this specific context, the textual information associated with each user is considered a `document`. For a word that is prevalent across all $N$ documents, the IDF is calculated as $log \frac{N}{N} = 0$, indicating that the IDF value for such a word tends to be small in the word vectors. This characteristic can potentially rectify the bias towards popular words.
\item \textbf{Pre-trained models}. \dg{} uses encoders pre-trained on a large and diverse corpus that covers many domains and languages, such as Sentence Transformers~\cite{reimers2019sentence} and CLIP. The resultant embeddings might also have learned some general representations for rarely used words, or at least some subword units that can approximate them. Consequently, the embeddings of sentences that incorporate these seldom-used words could still hold significance and prove beneficial for downstream tasks, such as the construction of an item-item graph in \dg{}. However, a potential limitation arises if the current pre-trained models solely depend on word frequency to generate sentence embeddings. This could lead to the production of generic or redundant summaries that fail to accurately reflect the users’ preferences.
\end{itemize}

\section{Conclusion}
In this paper, we propose \dg{} that is capable of encoding and disentangles both ratings and multimodal information with GCNs. 
By performing graph convolutions on a frozen item-item graph, \dg{} can learn the disentangled latent representations of items from their neighbors.
To enable the model interpretability, we project multimodal information into text leveraging pre-trained multimodal models and map user preferences with the words of their interacted items.
By regularizing the latent variables learned from user-item ratings and user textual preferences with mutual information maximization, \dg{} can interpret the decisions of users with words.
We evaluate the performance of our model on three real-world datasets against the state-of-the-art baselines.
The experimental results show that \dg{} outperforms the strongest baseline with significant gains.
Finally, we demonstrate the model interpretability by exemplifying an instance in a real-world dataset.

%


\begin{table*}[bt]
	\centering
	\def\arraystretch{0.85}
	\caption{Detailed multimodal information associated with the evaluated products of Amazon.}
	\label{tab:product_details}
	\begin{tabular}{p{0.1\textwidth} p{0.33\textwidth} p{0.5\textwidth}}
		\hline
		\textbf{Item ID} & \textbf{Title} & \textbf{Cover Image Hosted on Amazon} \\
		\hline 
		\multirow{3}{*}{B004M6UDC8} & Alpine Swiss Zippered Bifold Men's Wallet with Deluxe Credit Card Flip Pocket Genuine Lambskin Leather Comes in a Gift Bag-Black & \multirow{3}{*}{\url{http://ecx.images-amazon.com/images/I/41b--etdxtL._SX342_.jpg}} \\
		\hline 
		\multirow{2}{*}{B002YOMJPY} & Timberland Men's White Ledge Mid Waterproof Ankle Boot & \multirow{2}{*}{\url{http://ecx.images-amazon.com/images/I/41qzONzIe1L._SX395_.jpg}} \\
		\hline 
		B0007UMA3I & Mens Leather Zippered Credit Card Case & \url{http://ecx.images-amazon.com/images/I/41OlFMkIq1L._SX342_.jpg} \\ 
		\hline 
		\multirow{2}{*}{B004FG8BY2} & Invicta Men's 90242-001 Chronograph Black Dial Black Leather Dress Watch & \multirow{2}{*}{\url{http://ecx.images-amazon.com/images/I/51Qt3faT4-L._SY300_.jpg}} \\
		\hline 
		B005Z2LQ56 & Extra Warm Metro Fleece Ear Warmers & \url{http://ecx.images-amazon.com/images/I/31oIsqGtKnL._SX342_.jpg} \\
		\hline 
		\multirow{2}{*}{B003MYUQKA} & Invicta Men's 6977 Stainless Steel and Black Polyurethane Watch & \multirow{2}{*}{\url{http://ecx.images-amazon.com/images/I/51tAWqLFSWL._SY300_.jpg}} \\
		\hline
	\end{tabular}
\end{table*}

\section*{Acknowledgments}
This research is supported by Alibaba-NTU Singapore Joint Research Institute (JRI), Nanyang Technological University, Singapore.

{\appendices
\section{Detailed Descriptions of Interpreted Products}
\label{append:product}
We provide links to cover images of the interpreted products hosted on Amazon in Table~\ref{tab:product_details}.
These products are studied in Section \ref{sec:interp}.

\section{Implementation Details}
\label{append:imple}
We denote the numerical embeddings pre-trained from Sentence Transformer~\cite{reimers2019sentence} on items' textual contents as $\bm{X}_t$.
Each row in $\bm{X}_t$ denotes the embedding of an item in textual modality.
Analogously, the numerical embeddings in visual modality are denoted as $\bm{X}_v$.
We represent the user preference vectors generated with tf-idf as $\bm{R}^w$.
Table~\ref{tab:hy_mm} summarizes the hyperparameter settings and multimodal information used in each multimodal baseline. 
For each baseline's hyperparameters, we use their default values mentioned in the papers if not implemented with the official codebase. 

\begin{table}[bpt]
	\centering
		\def\arraystretch{.85}
	\caption{Implementation details of multimodal baselines. Grid searching is performed on each baseline to ensure optimal performance.}
	\label{tab:hy_mm}
	\begin{tabular}{p{0.07\textwidth} p{0.09\textwidth} p{0.24\textwidth}}
		\hline
		\multirow{2}{*}{\textbf{Model}} & \textbf{Multimodal Features} & \multirow{2}{*}{\textbf{Hyperparameters}} \\
		\hline 
		\multirow{2}{*}{VBPR} & \multirow{2}{*}{$\bm{X}_t$, $\bm{X}_v$} & regularization weight: \{2.0, 1.0, 0.1, 0.01, 0.001, 0.0001, 1e-05\}\\
		\hline 
		\multirow{4}{*}{MMGCN} & \multirow{4}{*}{$\bm{X}_t$, $\bm{X}_v$} & regularization weight: \{0, 0.00001, 0.0001, 0.001, 0.01, 0.1\}\\
		& & learning rate: \{0.0001, 0.0005, 0.001.0.005, 0.01\}\\
		\hline 
		\multirow{4}{*}{GRCN} & \multirow{4}{*}{$\bm{X}_t$, $\bm{X}_v$} & regularization weight: \{0.00001, 0.0001, 0.001, 0.01, 0.1\}\\
		& & learning rate: \{0.0001, 0.001, 0.01, 0.1, 1\}\\
		\hline 
		\multirow{4}{*}{DualGNN} & \multirow{4}{*}{$\bm{X}_t$, $\bm{X}_v$} & regularization weight: \{0.1, 0.01, 0.001, 0.0001, 0.00001\}\\
		& & learning rate: \{0.1, 0.01, 0.001, 0.0001, 0.00001\}\\
		\hline 
		\multirow{4}{*}{LATTICE} & \multirow{4}{*}{$\bm{X}_t$, $\bm{X}_v$} & regularization weight: \{0, $10^{-5}, 10^{-4}, 10^{-3}$\}\\
		& & learning rate: \{0.0001, 0.0005, 0.001, 0.005\}\\
		\hline 
		\multirow{4}{*}{MVGAE*} & \multirow{4}{*}{$\bm{X}_t$, $\bm{X}_v$} & regularization weight: \{0.0001, 0.001, 0.01, 0.1, 0\}\\
		& & learning rate: \{0.0001, 0.001, 0.01, 0.1\}\\
		\hline 
		\multirow{4}{*}{SLMRec} & \multirow{4}{*}{$\bm{X}_t$, $\bm{X}_v$} & learning rate and regularization weight: \{0.0001, 0.001, 0.01, 0.1\} \\
		& & $\tau_{ssl}, \tau$: \{0.1, 0.2, 0.5, 1.0\} \\
		& & $\alpha$: \{0.01, 0.05, 0.1, 0.5, 1.0\} \\
		\hline
		\multirow{3}{*}{BM3} & \multirow{3}{*}{$\bm{X}_t$, $\bm{X}_v$} & regularization weight: \{0.01, 0.1\} \\
		& & dropout: \{0.3, 0.5\} \\
		& & No. of GCN layers: \{1, 2\} \\
		\hline
		\multirow{2}{*}{DMRL*} & \multirow{2}{*}{$\bm{X}_t$, $\bm{X}_v$} & $\lambda{\theta}, \lambda_d$: \{$1e^{-5}, 1e^{-4}, \cdots, e$\} \\
		& & No. of the factors: \{1, 2, 4, 8\} \\
		\hline
		\multicolumn{3}{l}{\begin{tabular}{@{}l@{}l@{}} *In order to maintain consistency with other multimodal methods, DMRL \\does not utilize any review data. Additionally, we have consulted with the \\authors to ascertain the performance of MVGAE. \end{tabular}}%
	\end{tabular}
\end{table}
}

\bibliographystyle{IEEEtran}
\bibliography{reference}

\vfill

\end{document}